\documentclass[twocolumn,preprintnumbers,amsmath,amssymb]{revtex4}
\usepackage{dcolumn}% Align table columns on decimal point
\usepackage{amstext}
\usepackage{amsmath}
\usepackage{amssymb}
\usepackage[pdftex]{graphicx}
\usepackage{subfigure}
\usepackage[T1,OT1]{fontenc} 

\begin{document}
\title{Mechanism of membrane tube formation induced by adhesive nanocomponents}
\author{An{\fontencoding{T1}\selectfont\dj}ela \v{S}ari\'c and Angelo Cacciuto*}
\affiliation{Department of Chemistry, Columbia University\\ 3000 Broadway, MC 3123\\New York, NY 10027 }
\renewcommand{\today} 

\begin{abstract}
We report numerical simulations of membrane tubulation driven by large colloidal particles. Using Monte Carlo simulations we study how the process depends on particle size, concentration and binding strength, and present accurate free energy calculations to sort out how tube formation compares with the competing budding process. We find that tube formation is a result of the collective behavior of the particles adhering on the surface, and it occurs for binding strengths that are smaller than those required for budding. We also find that long linear aggregates of particles forming on the membrane surface act as nucleation seeds for tubulation by lowering the free energy barrier associated to the process. 
\end{abstract}
\maketitle

%\section{Introduction}
A key factor in cell trafficking and intercellular communication is the 
internalization of complex macromolecules. 
As large and charged biological cargo cannot directly cross the lipid bilayer that envelops the different compartments within eukaryotic cells, 
this process is usually accompanied by the formation of vesicular- and tubular-shaped membrane protrusions.    
The mechanism by which they develop can be extremely diverse ~\cite{camilli,basserau,helenius,lamb,schmidt,pagano}. It often
involves active processes requiring accessory factors, such as clathrin or caveolin protein coats, or motor-proteins and external forces. It can  
also develop as  a result of the self-assembly of  anchoring  proteins, such as BAR domain proteins~\cite{kozlov,prinz}, that impose a local curvature on the lipid bilayer. The physical mechanism driving protein aggregation in this case is fairly well understood within the framework of effective bending mediated Casimir forces\cite{pincus}. The size and shape of the resulting deformation is
determined  by  how the packing properties of the proteins couple to the elastic response of the membrane. 

Several endocytic pathways, however, are found to be triggered by the cargo itself~\cite{basserau,lamb,deserno, basserau1}. 
In some cases, such as HIV-1 ~\cite{rozycki}, the virus itself is formed on the membrane as its proteins
self-assemble inducing their own vesicular bud. The internalization is thus a consequence of cooperativity of many protein molecules. In this paper we are interested in passive internalization of preassembled viruses, virus-like particles and other colloidal particles. The main difference from the cases discussed above is that the interaction of a single colloidal particle  (typically one order of magnitude larger than a protein) with the lipid bilayer can induce its own invagination by wrapping its surface with the membrane. For instance, it has been shown that budding of preassembled alphaviruses and type-D retroviruses ~\cite{lamb,garoff}, as well as charged colloids ~\cite{granick}, can take place without the presence of external factors.
 
Although one might expect budding to be the main mechanism for internalization of large particles,
long tubular protrusions typically of one-particle diameter are often observed in viral or nanoparticle
internalization. Simian virus 40 (SV40), upon its binding to membrane receptors, is found to induce deep invagination and tubulation of both the plasma membrane and giant unilamellar vesicles (GUVs) ~\cite{SV40}. Its entry occurs via small, tight-fitting indentations and the resulting invaginations have the same size as the virus-particle diameter. Positively charged nanoparticles were also shown to spontaneously induce tubulation in supported ~\cite{orwar} and unsupported ~\cite{granick} giant unilamellar vesicles, suggesting the existence of a general physical mechanism of internalization, which is not exclusive for viruses and does not require assistance of membrane proteins. 

Understanding this phenomenon is of great importance for developing anti-viral strategies, but also because viral and virus-like particles, as well as artificial nanoparticles, are promising tools in gene-therapy and molecular medicine, for which control over their cellular uptake is essential. Despite the large body of work \cite{lipowsky1,deserno1,deserno2,balasz,may,chen,frenkel,zhang} on the particle budding problem,  most studies 
have focused on the interaction of a single particle with the membrane, and have completely missed tube formation, a  crucial component of the phenomenological behavior associated to particle internalization, that can only arise as a result of nontrivial cooperative behavior among many particles. Here we use computer simulations to investigate the physical mechanisms behind the occurrence of this process, and show how it depends on particle size, concentration and binding strength. While the phenomenon has been observed in several experiments, to the best of our knowledge, this paper presents the first theoretical study that addresses nanoparticle-driven tubulation, and  rationalizes its interplay with the particle budding process.   

Our system setup consists of $N_p$ particles, modeling colloidal viruses, virus-like particles or inorganic colloids, placed inside a vesicle of undeformed average radius $R$. 
Given the large size difference between the thickness of the vesicle and the nanoparticles considered in this study,
we model the vesicle using a simple one particle-thick solvent-free model consisting of $N$ spherical beads of diameter $\sigma$ connected by flexible links to form a triangulated network~\cite{nelson,ho,gompper} whose connectivity is dynamically 
rearranged to simulate the fluidity of the membrane. 
The membrane bending energy acts on neighboring triangles, and has the form 
\begin{equation}
E_{ij}=\frac{\kappa_b}{2}(1-{\bf n}_i\cdot {\bf n}_j)\,,
\end{equation}
where $\kappa_b$ is the bending rigidity, and the ${\bf n}_i$ and ${\bf n}_j$ are the normals of two triangles $i$ and $j$ sharing a common edge.  The cost associated with area changes is included via the energy term 
 $E_{\gamma}=\gamma A$, where $\gamma$ is the tension of the surface and $A$ is its total area.
The particles are represented as spheres of diameter $\sigma_{np}=Z\sigma$,
where $Z>1$ is a parameter used to control their size. Excluded volume between any two spheres in the system (particles and surface beads) is enforced with a hard-sphere potential. 
Finally, the colloid-to-membrane adhesion energy is modeled via
an additional power-law interaction between the particles and the surface beads defined as
\begin{equation}
V_{\rm {att}}(r)=-D_0 \left(\frac{\sigma_M}{r}\right)^{6}
\end{equation} 
and cut-off at $r_{\rm cut}=1.5\sigma_M$, with $\sigma_M=(\sigma+\sigma_{np})/2$. $D_0$ is thus the membrane-particle binding constant. This potential is quite generic and is typically employed to describe short-range attractions, such as ligand-receptor or van der Waals interactions. The system is equilibrated using the Monte Carlo simulations in the NVT ensemble and most of our data are obtained at $\kappa_b=5 k_{\rm{B}}T$, $\gamma=1 k_{\rm{B}}T/\sigma^2$ and  $Z=2,3,4,6 $ or $8$.\\

To qualitatively understand the interaction between a single particle and a membrane, consider a particle of radius $R_p$ having a spherical cap of height $h$ and area 
$S_{\rm cap}=2\pi R_ph$ in contact with a membrane. The elastic costs associated with this configuration are $\frac{k_b}{R_p^2}S_{\rm cap}$ and  $\gamma \pi h^2$, where the first term accounts for the bending and the second for the surface energy of the membrane.
The free energy gain due to the binding energy between particle and membrane scales as $-D_0S_{\rm cap}$.
A balance of these terms leads to an equilibrium  particle-membrane indentation, $h$, and to a particle  coverage
$\chi_T\equiv S_{\rm cap}/(4\pi R_p^2)=\frac{D_0-k_b/(2R_p^2)}{2\gamma}$. This suggests that the particle will bud off the membrane as soon as  $D_0 \geq k_b/(2R_p^2)$. Although more sophisticated calculations have been put forward 
to understand the nature of this transition ~\cite{lipowsky,chan,suresh,deserno1}, the main result is that for a 
given binding constant $D_0$, budding is easier for large particles.
This scaling argument gives a simple explanation of  why this process is likely for colloidal particles and preassembled viruses, but not for single proteins and  small macromolecules, and provides us with an intuitive framework from which to begin our study.

\begin{figure*}[n]
\center
\includegraphics[width=100mm]{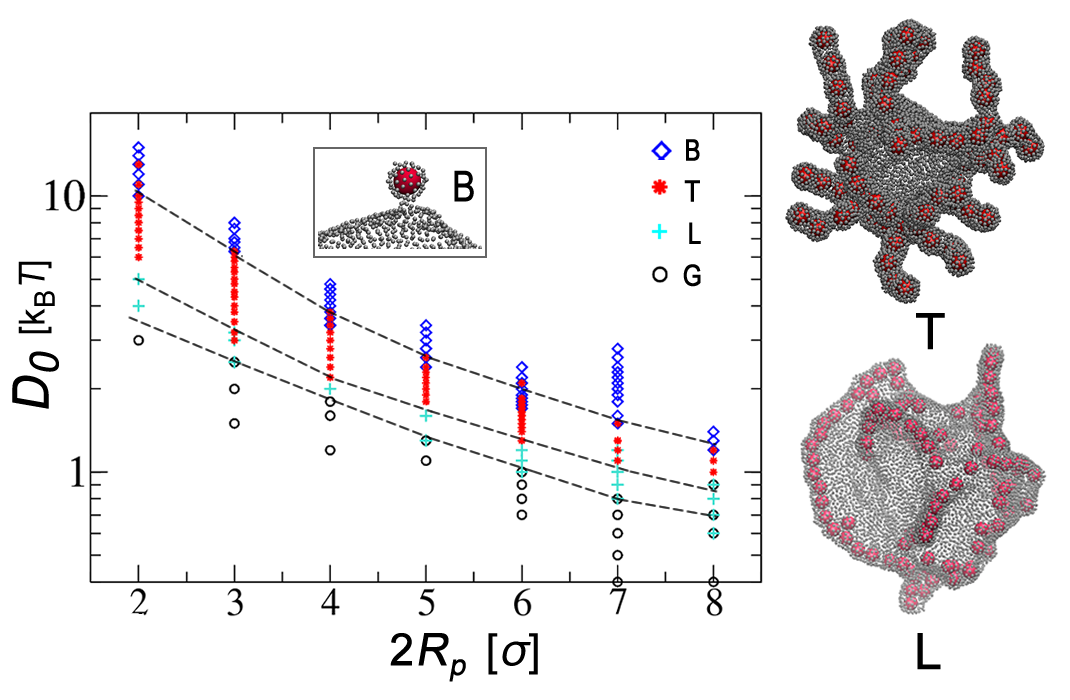}
\caption{ Left panel: $D_0$-$R_p$ phase diagram of the membrane aggregates and protrusions induced by colloidal particles. Right panel: Snapshots of the linear (\textbf{L}) and tubular (\textbf{T}) phases. The inset shows a typical single-particle bud conformation (\textbf{B}) that occurs at large $D_0$. The bottom region of the phase diagram is the gaseous phase (\textbf{G}). The radius of the membrane is $R=30\sigma$ and the particle surface fraction is kept constant at 0.15.}
\label{fig1}
\end{figure*}

We begin our analysis by computing a diagram that indicates, for a given value of $D_0$ and $R_p$, the phenomenological behavior of the particle-membrane coupled system at a constant particle concentration.
The results are shown  in Fig.~\ref{fig1}. 
%The important length-scale for the membrane deformation can be constructed out of its specific parameters as $R_0\sim\sqrt{\kappa_b/%\gamma}$. We keep this length-scale constant, but change the relative ratio $R_p/R_0$.
For small values of $D_0$, the overall shape of the membrane is unchanged while the particles, barely adhering to it,
freely diffuse over its surface as a low-density two-dimensional gas (\textbf{G}).  
Increasing $D_0$, we find that the nanoparticles organize into linear aggregates (\textbf{L}). 
This phase develops due to effective interactions between the particles driven by the membrane's need to minimize its elastic energy while 
maximizing its binding surface to the particles (see ~\cite{cacciuto, safinya} for a detailed analysis of this phase and its experimental observation).  
%Surprisingly, we find this phase to be a precursor for the next phase transformation that takes place 
Upon further increase in $D_0$,  spontaneous formation of tubular protrusions (\textbf{T}) takes place. 
This region of the diagram is characterized by nanoparticles tightly and linearly
 packed into tubular structures extruding out of the membrane core. The radius of the tubes equals the diameter of the particles.
% gaining twice as much membrane contact than in the \textbf{L} phase. 
This behavior is in agreement with the SV40-induced membrane invaginations~\cite{SV40}, where one-particle-wide tubes were also observed, but tubulation failed to occur if the adhering viruses were unable to form a sufficient number of interactions with the membrane binding sites.
Further increase in $D_0$ causes nanoparticles to promptly adhere to the membrane and become completely enveloped into a bud 
before any significant particle diffusion can occur.  The \textbf{T}-\textbf{B} transition is not abrupt, and a mixture of both ``corrugated'' tubes and single-particle buds is found in the borderline area between the two phases. Although in our model buds cannot physically detach from the membrane, they are easily identifiable by their complete surface-coverage and the characteristic sharp membrane neck shape. A single particle bud is shown in the inset of Fig.~\ref{fig1}.

The most important message arising from our analysis is that tubulation develops as a result of the interaction of many particles and should be expected for intermediate binding constants.
Such a behavior occurs for all particle sizes considered in this study, indicating what sets the 
tube size is the particle diameter and not the natural length-scale associated to membrane tube formation, $R_0=\sqrt{\kappa_b/(2\gamma)}$, obtained by pulling experiments~\cite{dogterom,derenyi}. 
Moreover, preassembly of nanoparticles into linear aggregates seems to greatly facilitate the formation of long tubes.     

\begin{figure*}
\center
\subfigure[]{\label{fig2a}\includegraphics[width=59mm]{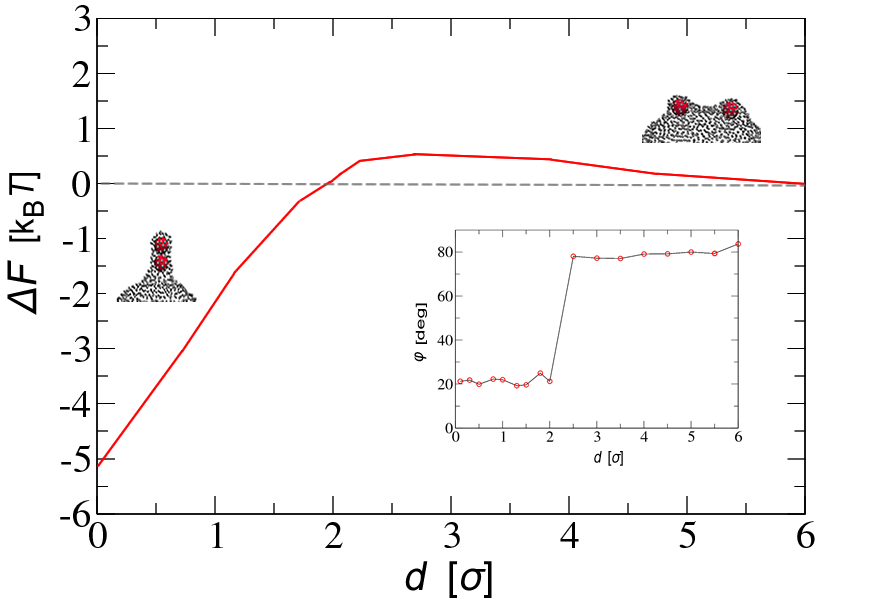}}
\subfigure[]{\label{fig2b}\includegraphics[width=61mm]{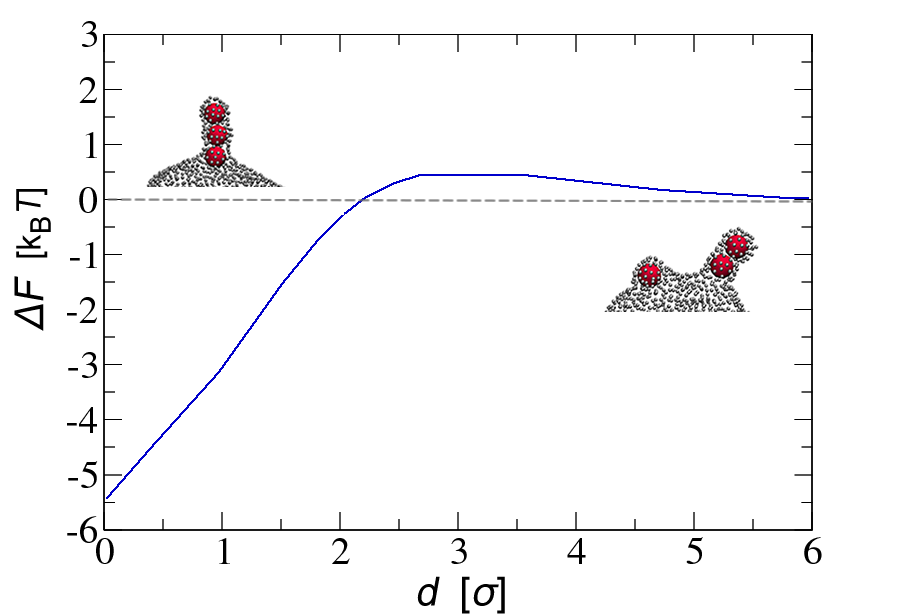}}
\caption{ Tube formation and growth. \textbf{(a)} Free-energy as a function of separation of two membrane-bound particles. The inset shows the orientation of the dimer with respect to the membrane surface and indicates the distance at which the tubulation occurs. Here, $\varphi$ is defined as the angle formed by the vector connecting the centers of the two particles and that connecting the center of the membrane to the particles' midpoint. \textbf{(b)} Free-energy as a function of separation of a two-particle tube and a single membrane-bound particle. In both cases $Rp=4$, $R=15\sigma$ and $D_0=2.6k_{\rm{B}}T$.}
\end{figure*}

To obtain more physical insight into the mechanism by which tubular protrusions form, we considered a series free energy calculations. 
First, we measure the effective interaction between two colloids adhering to the membrane in the \textbf{T}-region of the phase diagram. Using the umbrella sampling method~\cite{valleau}, we compute the probability that the two particles are at any given separation $d$ from each other and estimate the free energy difference $\Delta F=F(d)-F(\infty)$. Fig.~\ref{fig2a} shows $\Delta F$ as a function of $d$, while the inset monitors how the orientation of the dimer with respect to the membrane surface, $\varphi$, depends on the same variable. This result is quite revealing; the elastic cost required to bring together two large membrane deformations, responsible for the weak mid-range repulsion, is replaced by a large energy gain when the particles are in contact. The corresponding configuration is characterized by two particles contained within a membrane tube oriented perpendicularly to the membrane surface. As we have not imposed any constraint on the values of $\varphi$, this is clear evidence, at least at the two-particle level, that 
in this region of the phase diagram, tube formation is more favorable than budding. 
%The well depth basically gives us the free energy gain of tubulation of two particles, which is about $5k_{\rm{B}}T$ just above the \textbf{L}-\textbf{G} boundary and increases with increasing $D_0$ (results not shown).

Using the same procedure, we can also measure the free energy as a function of separation between a two-particle-tube and a third isolated particle. Our data, shown in Fig.~\ref{fig2b}, tells us that the lowest free energy is again achieved when the three particles are in contact in a tubular formation. This very important result indicates that tubes and free  particles bound to the membrane attract each other, and once a tube is formed, its growth by particle addition drives the system towards a lower free energy. In both cases, the extent of the repulsion and attraction is dependent on the specific region of the phase diagram they are computed at. The characteristic energy barrier at mid-range distance becomes more significant as $D_0$ increases and the system crosses over to the budding regime, implying that for large $D_0$ particle aggregation becomes rare, making budding the most likely barrier-crossing mechanism. This is a kinetically dominated regime: in fact, once the budding threshold has been overcome, particles would leave the membrane before having the time to aggregate.

\begin{figure*}
\center
\includegraphics[width=70mm]{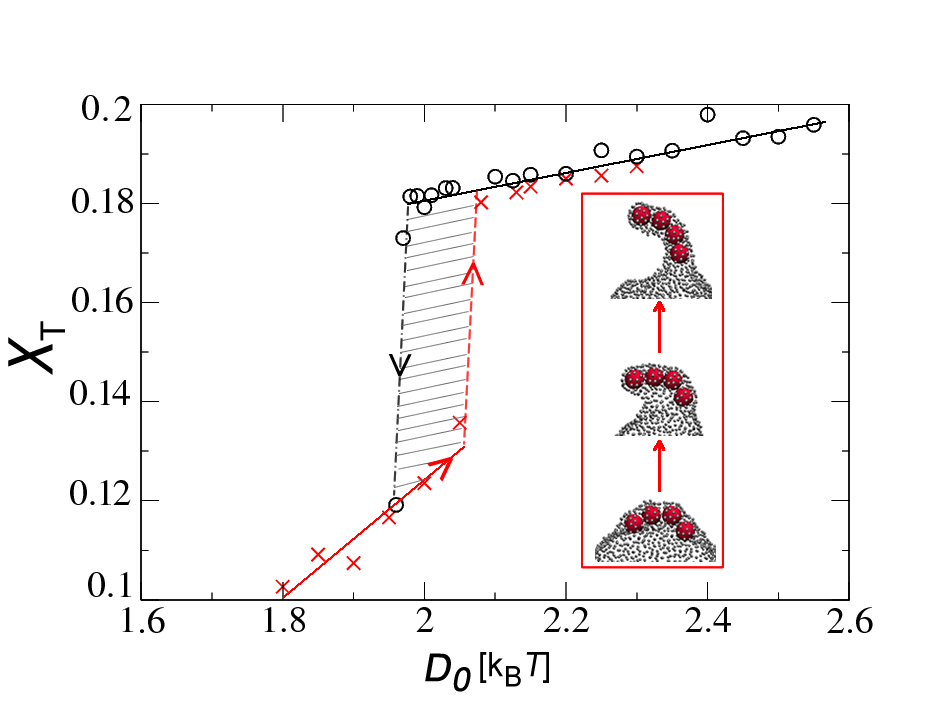}
\caption{Hysteresis associated with the tubulation of a linear aggregate, in terms of the surface coverage $\chi_T$ and $D_0$, for the extrusion of a four-particle-long aggregate. $\chi_T$ is computed as the ratio between the number of membrane beads in contact with the particles and the same number when the surface completely envelops the particles. The red crosses show the results of simulations that start form a linear aggregate, while the black circles show simulations that start from a tube.  Here $Rp=4\sigma$ and $R=15\sigma$}. 
\label{fig3}
\end{figure*}

\begin{figure*}
\center
\includegraphics[width=70mm]{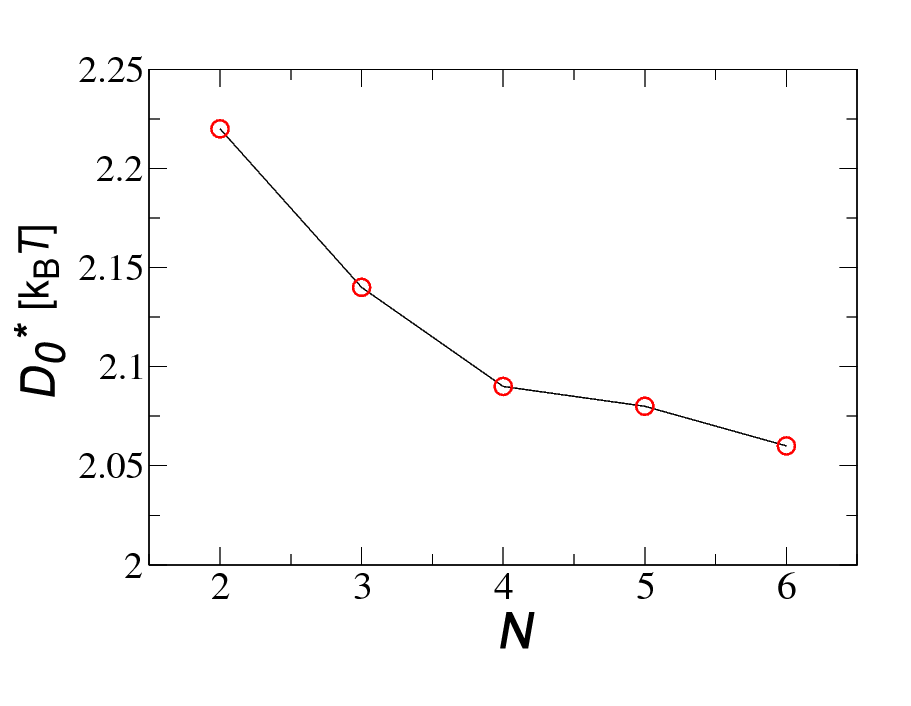}
\caption{Onset binding constant for tubulation $D^*_0$ as a function of the length of the preformed linear nucleation cluster; $Rp=4\sigma$ and $R=15\sigma$.}
\label{fig4}
\end{figure*}

Interestingly, in most of our simulations in the \textbf{T} phase we observe that tube formation if often preceeded, in particular at higher particle densities, by the formation of long linear aggregates that eventually extrude from the membrane via a tilting mechanism illustrated in Fig.~\ref{fig3}. This two-step process becomes more significant as we move closer to the \textbf{L}-\textbf{T} boundary, suggesting that these aggregates function as nucleation seeds promoting the transition. To support this idea we perform two sets of simulations: in the first set we start from a connected four-particle-long linear aggregate, and measure its surface coverage $\chi_T$ as a function of $D_0$ until a tube is formed, in the second set we start from an already tubulated structure and we decrease $D_0$ until the tubule retracts. As shown in Fig.~\ref{fig3}, tubulation is accompanied by a sudden jump in the particle coverage $\chi_T$ (and consequentially in the binding energy), indicating the presence of a free energy barrier between the two states that needs to be crossed for the linear aggregates to protrude out of the membrane. This result is consistent with previous force-extension calculations and experiments on GUVs~\cite{dogterom}, that also indicated tube formation to be a first order transition.
Finally, we measured the onset value $D^*_0$ at which a preformed linear aggregate forms a tube as a function of its size, at a fixed particle radius. 
A weak but clearly inverse dependency is found, shown in Fig.~\ref{fig4}, and supports the idea that the free energy cost for tubulation from the \textbf{L} phase does indeed decrease monotonically with the size of the aggregates which therefore act as nucleation seeds for the transition. It should be stressed that because the probability of forming linear aggregates increases with particle concentration, it is logical to expect tubulation to be more likely to occur in denser systems. This is indeed what we find in our study.

%Interestingly, if we interpret particle-driven tubulation as budding of a cluster of particles, by ideally extrapolating the data in Fig.~\ref{fig4} to $N=1$ we have further evidence in favor of tubulation versus budding as the main mechanism of particle internalization in this region of the phase diagram. 
 
We have shown that for a broad range of binding energies, tube formation and not membrane budding is the
main mechanism leading to internalization of sufficiently large particles.
Nowhere in our simulations have we observed formation of membrane tubes of radius larger than one particle diameter;
however, these may develop as a result of direct particle-particle interactions or nontrivial long-range electrostatic effects~\cite{granick} not
included in our study.  It should be emphasized that our results should hold as long as the particle size is sufficiently large so that the molecular details of the membrane can be ignored. Although the elastic constants of our  model were selected in a range relevant to biological processes
and we only considered two vesicle radii, we do not expect the process to be extremely sensitive to these parameters.
It would be useful to carry out a surface minimization analysis of this problem as it was performed for force driven tubulation~\cite{derenyi},
and we really hope that our work will stimulate further  experimental work in this direction.

\section*{ACKNOWLEDGMENTS}
This work was supported by the National Science Foundation under Career Grant No. DMR-0846426.

\vspace{50mm}

\end{document}